\documentclass[11pt]{article}
\usepackage{moriond,epsfig}

\def\be{\begin{equation}}
\def\ee{\end{equation}}
\def\bea{\begin{eqnarray}}
\def\eea{\end{eqnarray}}

\newcommand{\half}{\raisebox{1pt}{$\scriptstyle \frac{1}{2}$}}
\newcommand{\quarter}{\raisebox{1pt}{$\scriptstyle \frac{1}{4}$}}

\begin{document}
\begin{flushright}
MC-TH-2003-3
\end{flushright}
\vspace*{2cm}
\title{Two novel possibilities in Higgs phenomenology}

\author{J.R. Forshaw}

\address{Department of Physics and Astronomy, University of Manchester, 
Oxford Road,\\
Manchester M13 9PL, England.}

\maketitle\abstracts{
We discuss two novel possibilities in Higgs physics. The first is that, by 
adding a real Higgs triplet to the Standard Model, it is possible for 
the lightest Higgs boson to be as heavy as 500~GeV without any fine tuning. 
The second, somewhat orthogonal, possibility concerns the MSSM with explicit 
CP violation. This model is
known to permit a light Higgs boson with mass below 50~GeV which may 
have avoided detection at LEP and may also avoid detection at the Tevatron 
and LHC. We suggest that diffraction may provide the key to excluding or 
observing such a Higgs. 
}

\section{Introduction}
Precise data from LEP, SLC and Tevatron imply a light Higgs boson when 
interpreted within the Standard Model~\cite{lepewwg}, i.e. 
$m_h = 81^{+52}_{-33}$~GeV, and similar conclusions 
also apply in the simplest
supersymmetric extensions. It is natural to ask how general this situation
is. In this talk, we focus upon two quite different extensions to the
Standard Model which avoid the more usual constraints on the Higgs mass.
In the first case, we consider adding an additional real Higgs triplet to
the Standard Model with the effect that the lightest Higgs could be as
heavy as 500~GeV. The model is notable in that there is no severe fine tuning
and no problem with unwanted phenomenology. Moreover, the argument allowing
a heavier Higgs is essentially tree-level. In Section 3, we turn to the
MSSM with explicit CP violation. CP violation in the Higgs sector can
lead to a situation where the lightest Higgs is
lighter even than the direct search limit for a Standard Model Higgs. We
suggest that diffractive scattering may provide a means for discovering or 
excluding such a Higgs.

\section{Triplet Higgs}
The model~\cite{triplet,frw} consists of adding a real hypercharge zero 
Higgs triplet ($H$) to the Standard Model doublet ($\Phi$), i.e.
\bea
  \mathcal{L}(\Phi, H)  &=&  (D_\mu {}\Phi)^{\dagger }(D^{\mu }\Phi) \; + \; \half (D_\mu {}H)^{\dagger }(D^{\mu }H)
\; - \; V(\Phi, H), \nonumber \\
 V(\Phi, H) & =& \mu_1^{2} \, \Phi^{\dagger }\Phi \; + \; \half \mu_2^{2} \,
H^{\dagger }H \nonumber \\
 & +& \lambda_1 \, (\Phi^{\dagger }\Phi)^{2} \; + \; \quarter \lambda_2 \, (H^{\dagger }H)^{2} \nonumber \\
 &+& \half \lambda_3 \, (\Phi^{\dagger }\Phi)(H^{\dagger }H) \; + \; \lambda_4
\, v \, H_i \, \Phi^{\dagger }\sigma^{i} \Phi. \eea

Defining 
$\tan \beta$ to be the ratio of the triplet to doublet vacuum expectation
values then, for non-zero $\beta$, this model violates custodial symmetry, i.e.
\be
\rho \; \equiv \; \frac{m^{2}_{W}}{m^{2}_{Z} c^{2}_{W}} \; = \; 
\frac{1}{\cos^{2}\beta} \neq 1. \label{rho} 
\ee
Usually this tree-level deviation of $\rho$ from unity is regarded as
an unpallatable feature of the model. However 
it is precisely this symmetry breaking which allows the lightest 
Higgs to be much heavier than in the 
Standard Model~\cite{frw}. 
By giving the triplet a non-zero vacuum expectation value, 
one is in effect making a positive tree-level contribution to the 
$T$-parameter, and this is enough to allow a heavier Higgs.

\begin{figure}
\begin{center}
\includegraphics[clip=true,width=.35\textwidth]{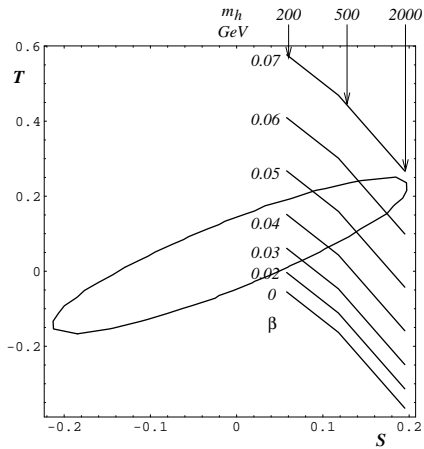}
\end{center}
\caption[]{Ellipse encloses the region allowed by data. Curves show results 
in the Triplet Model for various values of 
\protect\( \beta \protect \) and various doublet Higgs masses. \protect\(
\beta = 0 \protect \) is the Standard Model curve and the reference Higgs
mass is 100~GeV. Figure from Forshaw, Ross and White~\cite{frw}.}
\label{fig:triplet}
\end{figure}

This positive contribution to $T$ can be traced back to the use of the
muon decay constant in the precision electroweak fits. Equation (\ref{rho})
effectively induces a small shift in the $W$ mass, and this feeds back,
via $G_\mu$, into the measured observables. The
correction can be represented as a positive contribution to the oblique
parameter $T$, i.e. $\alpha \Delta T \approx \beta^2$. 
In Figure \ref{fig:triplet} we show the familiar ellipse in the $S-T$
plane. The interior of the ellipse is allowed by the precision data and
the curve corresponding to $\beta=0$ is the Standard Model result for
various Higgs masses, clearly in this case the Higgs should be light. 
For non-zero $\beta$, the Standard Model curve is shifted upwards thereby
allowing the lightest Higgs to be much heavier.

Some additional comments are in order. Firstly, since the triplet
has no hypercharge it does not contribute to the oblique parameter $S$.
Secondly, the genuine quantum corrections to $T$ are naturally small
since, for small mixing $\beta$, the charged and neutral members of the
triplet have small mass splitting. In other words, quantum corrections
are, to a first approximation, negligible. Note also that $\beta$ does not
need to be particularly small in order to accommodate the data. 
Thirdly, we mention that a study of the mass spectrum in the triplet model 
and the RG flow of the scalar couplings has recently been 
completed~\cite{fsw}. Finally, one should note that there are other 
ways to avoid 
a light Higgs~\cite{peskin}. In the case that there are no new light
particles seen at the LHC, decisive information could well come
from even more precise measurements of the $Z$ boson~\cite{gigaz}.

We now leave behind the triplet model and shift focus to the opposite
extreme of a Higgs which may be even lighter than suggested by the
Standard Model.

\section{CP violating light Higgs}

\begin{figure}
\begin{center}
\includegraphics[clip=true,width=.45\textwidth]{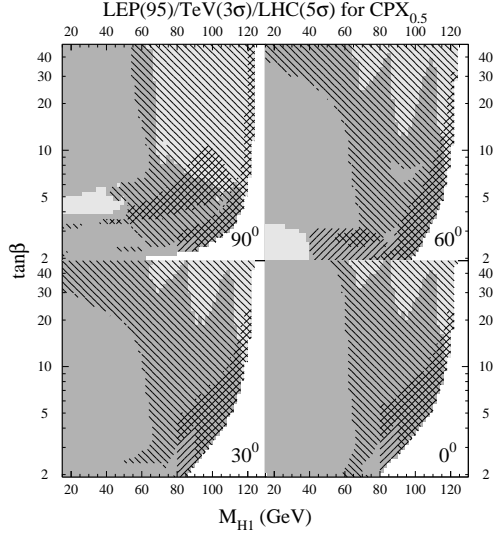}
\end{center}
\caption[]{Approximate Tevatron/LHC discovery and 
LEP exclusion limits in the $M_{H_1}$--$\tan\beta$ plane for the CPX 
scenario with both phases set to: (a) $90^\circ$, (b) $60^\circ$,
(c) $30^\circ$, and (d) $0^\circ$.
The reach of the Tevatron $W/ZH_i(\to 
b\bar b)$ search is shown as $45^\circ$ lines and that of the combined
LHC search channels as $135^\circ$ lines.The combined LEP 
exclusion is shown in medium gray, superimposed on the
theoretically allowed region in light grey. Figure from 
Carena et al.~\cite{exclude}}
\label{fig:cpx}
\end{figure}

The possibility of CP violation in the Higgs sector of the MSSM 
occurs quite naturally as a result of radiative effects induced by 
explicit CP violation in the third generation of squarks ~\cite{CPV}.
Moreover, it has been established that the lightest Higgs boson
could be much lighter than the direct search limit, obtained assuming the
Standard Model, due to a severely weakened coupling to the $Z$ 
boson. We refer to Carena et al and references therein for more 
details~\cite{exclude}. 

This is illustrated in Figure \ref{fig:cpx} where
we are particularly interested in the upper two plots which have large
CP violating phases. In these plots one should focus upon the unexcluded 
windows (light grey) 
at low to intermediate $\tan \beta$~\footnote{Note that $\beta$ here is
not the same as that in Section 2.} and $m_h < 50$~GeV. Note also
that it may be difficult to exclude these regions using
conventional search channels at either the Tevatron or LHC~\cite{chl,exclude}.

\begin{figure}
\begin{center}
\includegraphics[clip=true,width=.23\textwidth]{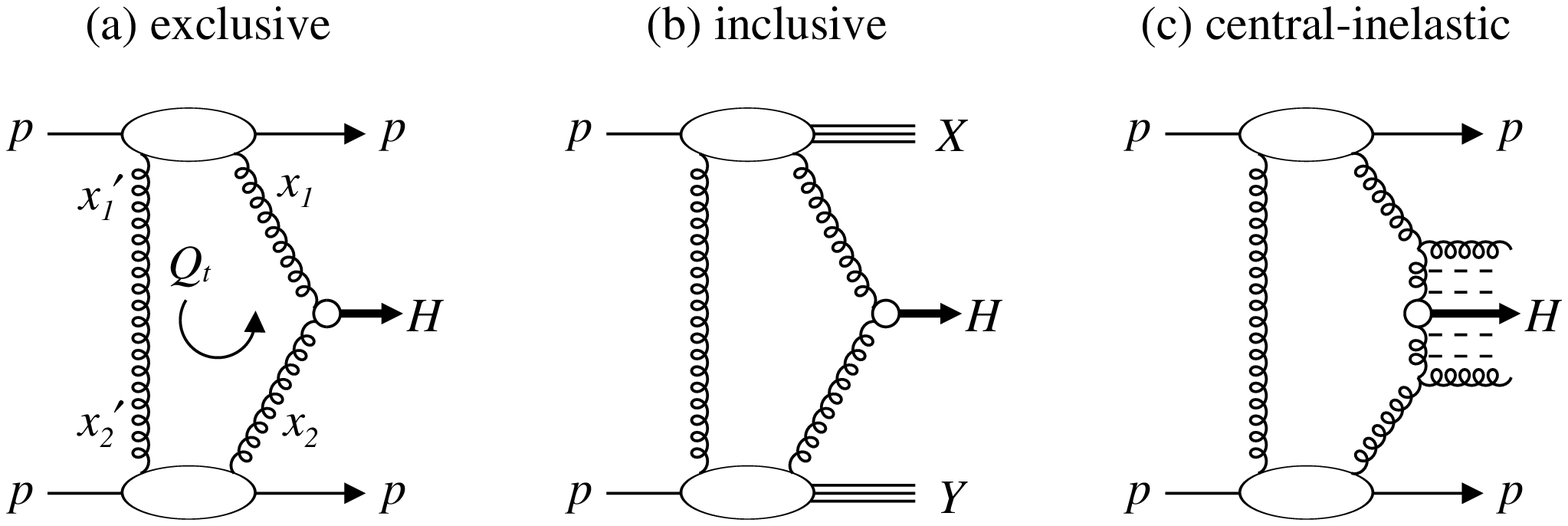}
\end{center}
\caption[]{Diagram for exclusive diffractive Higgs production. 
Figure from Khoze, Martin and Ryskin.~\cite{kmr}}
\label{fig:kmr}
\end{figure}

Fortunately, it is possible that such light Higgs bosons may be
visible in diffractive scattering events at hadron colliders via the
process $p+p \to p+H+p$, illustrated in Figure \ref{fig:kmr}. The Higgs
boson is produced centrally and the incoming hadrons remain intact,
deflecting through small angles. In our
calculation of the rate, we follow
closely the approach of Khoze, Martin and Ryskin~\cite{kmr};
more details can be found in our paper~\cite{cflmp}. We would like
to stress that a higher statistics measurement of the central production 
of jets at the Tevatron than has been obtained so far~\cite{CDF} should 
provide a good test of the underlying theoretical 
framework. Such a measurement should be possible in the near future.

\begin{figure}
\begin{center}
\includegraphics[clip=true,width=.65\textwidth]{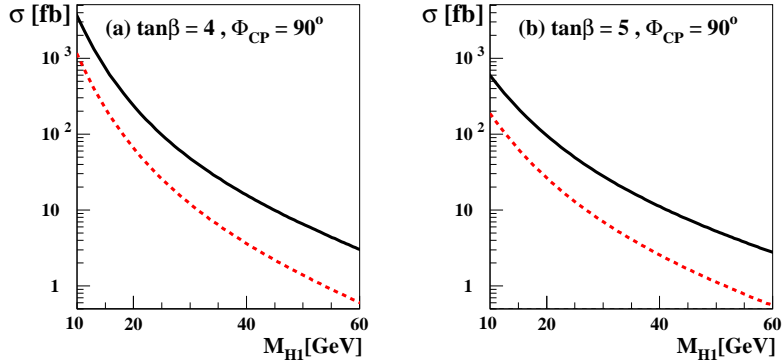}
\end{center}
\caption[]{Cross-section for diffractive Higgs production in the case
of a light CP violating MSSM Higgs as a function of the lightest Higgs 
mass. Solid line is for LHC energies and
the dotted line is for the Tevatron. Figure from Cox et al.~\cite{cflmp}}
\label{fig:xsecn}
\end{figure}

In Figure \ref{fig:xsecn} we show the resulting total cross-section for
$p+p \to p+H+p$ as a function of Higgs mass at both Tevatron and LHC
energies. The SUSY parameters are chosen to cover the unexcluded windows 
of Figure \ref{fig:cpx}. Note that for masses below 30~GeV, there may even be
sufficient rate to make exploration at the Tevatron a possibility. Certainly
these light Higgses ought to be produced in abundance at the LHC.

Although we have not yet performed a detailed study of the background, there
are good reasons to expect that it should be possible to isolate these 
diffractive events. The light Higgses are expected to decay
to $b$-quarks and the corresponding QCD background will be suppressed by
a selection rule which favours the production of $0^+$ states in 
diffraction (although this rule will become less effective for lower
Higgs masses). Moreover, a resolution of order 1~GeV
on the mass of the central system may be possible at the LHC with the
installation of suitable final state proton detectors \cite{deroeck}.

In summary, it may well be that central production will provide a unique and
valuable tool to complement more traditional search strategies for new
physics. To exploit this opportunity requires that suitable detectors be
installed to measure the momenta of the final state protons.

\section*{Acknowledgments}
Many thanks to the organizers of the Moriond conference for a very 
pleasant and stimulating week. Thanks also to my collaborators:
Brian Cox, Jae Sik Lee, James Monk, Apostolos Pilaftsis, Douglas Ross,
Agustin Sabio Vera and Ben White. This work has been supported in part
by the UK Particle Physics and Astronomy Research Council.

\end{document}